\documentclass{elsart}
\usepackage[dvips]{epsfig}
\newcommand{\AmS}{{\protect\the\textfont2
  A\kern-.1667em\lower.5ex\hbox{M}\kern-.125emS}}

\begin{document}
\begin{frontmatter}
\title{A monitor of beam polarization profiles for the TRIUMF parity 
experiment\thanksref{money} }

\author[CMU]{A.R.~Berdoz},
\author[Man]{J.~Birchall},
\author[Man]{J.B.~Bland},
\author[LANL]{J.D.~Bowman},
\author[Man]{J.R.~Campbell},
\author[Alta]{G.H. Coombes},
\author[TRI,Man]{C.A.~Davis\thanksref{correspond}},
\author[Alta]{P.W.~Green},
\author[Man]{A.A.~Hamian},
\author[INS]{Y.~Kuznetsov\thanksref{deceased}},
\author[Man]{L.~Lee},
\author[TRI]{C.D.P.~Levy},
\author[LANL]{R.E.~Mischke},
\author[Man]{S.A.~Page},
\author[Man]{W.D.~Ram\-say},
\author[Man]{S.D.~Reitzner},
\author[TRI]{T.~Ries},
\author[Alta]{G.~Roy},
\author[Man]{A.M.~Sekulovich},
\author[Alta]{J.~Soukup},
\author[Alta]{T.~Stocki},
\author[Man]{V.~Sum},
\author[INS]{N.A.~Titov},
\author[Man]{W.T.H.~van~Oers},
\author[Man]{R.J.~Woo},
\author[INS]{A.N.~Zelenski}
\address[Man]{University of Manitoba, Winnipeg, MB }
\address[LANL]{Los Alamos National Laboratory, Los Alamos, NM }
\address[Alta]{University of Alberta, Edmonton, AB }
\address[TRI]{TRIUMF, Vancouver, BC }
\address[INS]{Institute for Nuclear Research, Academy of Sciences, Moscow, Russia }
\address[CMU]{Carnegie Mellon University, Pittsburgh, PA }
\thanks[money]{Work supported in part by a grant from the Natural
Sciences and Engineering Research Council of Canada}
\thanks[correspond]{Corresponding author.  TRIUMF, 4004 Wesbrook Mall,
         Vancouver, B.C., Canada V6T 2A3; Tel.:1-604-222-1047, loc. 6316;
         fax: 1-604-222-1074 \\
      {\em E-mail address:} cymru@triumf.ca}
\thanks[deceased]{Deceased}

\begin{abstract}
TRIUMF experiment E497 is a study of parity violation in {\it pp}
scattering at an energy where the leading term in the analyzing
power is expected to vanish,
thus measuring a unique combination of weak-interaction flavour
conserving terms.  It is desired to reach a level of sensitivity
of $2 \times 10^{-8}$ in both statistical and systematic errors.
The leading systematic errors depend on transverse polarization
components and, at least, the first moment of transverse polarization.
A novel polarimeter that measures profiles of both 
transverse components of polarization
as a function of position is described.
\end{abstract}
\end{frontmatter}

\section{Introduction}
Several measurements of the parity violating component in the
nucleon-nucleon interaction have been reported over the years
\cite{kis:87} \cite{bal:84} \cite{eve:91} \cite{yua:86} \cite{pot:74}, 
achieving greater precision over time.  Such an
experiment that aims to measure longitudinal analyzing power,
$A_{z}$, to a precision of $ \pm 2 \times 10^{-8}$ 
(in statistics and systematics) is underway     
at TRIUMF \cite{E497}.  Other experimenters have measured the
same quantity with protons incident on light nuclei \cite{har:85} \cite{bow:75}.
The TRIUMF experiment is unique in that it seeks to measure the
parity-violating effect at an energy, 221 MeV, where the leading term
(which dominates below 100 MeV), $A_{z}(^{1}S_{0}-^{3}P_{0})$,
is zero (averaged over the acceptance of the detector), thus
observing the $A_{z}(^{3}P_{2}-^{1}D_{2})$ term as the dominant
component \cite{birch:95}.  The difference in these terms is that they
are dependent on different combinations of the weak meson couplings
\cite{birch:95}.  In addition, another experiment is planned
at 450 MeV at TRIUMF \cite{birch:II}.  

Initially, it was recognized that residual
transverse components of polarization which changed sign with
the longitudinal component of polarization gave rise to a systematic
error if the detection system was asymmetric or if the
incident proton beam were off the symmetry axis \cite{sim:72}.
Later, it was recognized that even if the transverse polarization
components of a finite-sized beam averaged to zero, an inhomogeneous
distribution of the transverse polarization over the beam profile
could result in a significant contribution to the measured $A_{z}$
\cite{sim:80} \cite{nag:78}.

The ETH-SIN-Z\"{u}rich-Karlsruhe-Wisconsin group \cite{kis:87} 
\cite{bal:84} describe
a beam intensity/polarization profile monitor \cite{hae:79}
that operates with two wheels ({\it x} and {\it y}) each driving
two graphite targets through their 50 MeV proton beam.  Protons
scattered at $51^{ \circ }$, near the maximum of the $^{12}C(p,p)^{12}C$
analyzing power, were observed in four scintillators left, right,
bottom and top, and timed with respect to a reference on the wheel,
i.e. the position of the target for that scattered proton.  The data,
along with information about the spin state, were read into a series
of spectra from which intensity and polarization profiles could be
deduced.  They used two such devices in their beamline.
This polarimeter is also described modified for use at lower energies
\cite{vua:86}.  The targets used were much thinner and they adapted
a multi-channel scaler and multi-channel analyzer to record and store
the profiles.

The Bonn group \cite{eve:91} describe a beam profile scanner that measures 
polarization \cite{chl:87} by physically moving a polarimeter (one for
vertical and one for horizontal profile) with
a thin graphite target through the beam.  The target is optimized
to allow passage of one target at a time through the beam while data collection
is enabled.  They detected protons scattered at $48 ^{ \circ }$.
An elastically scattered proton      %How do they know if the proton is e.s.?
in any of four Si detectors generated a sampling of an ADC that read a voltage
picked up from a linear potentiometer related to the device's 
position.  Again, position dependent spectra were generated from which
intensity and polarization profiles can be deduced.  They also used
two such devices in their beamline.

The Los Alamos-Illinois group \cite{yua:86} \cite{har:85} describe
a simple scanning target used in a conventional polarimeter.

\section{Specifications and design}
Requirements of the TRIUMF Parity experiment are that it be able to 
intimately (i.e., within the data collection cycles of the experiment,
each cycle being eight periods of 25 ms duration) monitor the profile
of the transverse polarization components as a function of position
($P_{y}(x)$ and $P_{x}(y)$), and that one be able to determine the corrections
derived therefrom to $A_{z}$ to a level at or below $ \pm 6 \times 10^{-9}$
over the whole data collection period (several hundred hours not counting
calibrations and other overhead).
These quantities are the average transverse components of polarization
which given an effect proportional to displacement from the detector
symmetry axis, $<x><P_{y}>$ and $<y><P_{x}>$, and the intrinsic first
moments of polarization, $<xP_{y}>$ and $<yP_{x}>$, which will contribute
even if the proton beam is perfectly aligned with the apparatus.
Higher order terms and in-plane terms (such as $<xP_{x}>$ and $<yP_{y}>$)
should be negligible \cite{sim:80} \cite{bal:84}.

Typical run time conditions keep transverse components of polarization
under $ \pm 1 \% $ per data run (typically one hour).  Intrinsic first 
moments of transverse polarization are typically within $ \pm 25 \mu \mbox{m}$
per run.

As mentioned above, some researchers \cite{hae:79} \cite{chl:87} \cite{vua:86} 
use graphite targets which have
relatively high counting rates and high analyzing powers.
However, the angular dependence of both cross section and analyzing power
and the contribution of inelastic scattering, especially at higher energies, 
make this undesirable in the present case.
By moving the polarimeter detectors rigidly with the target, Chlebek et al.
\cite{chl:87} avoid position correlated acceptance problems.  Such a 
device had initially been 
considered \cite{birch:88} but abandoned when it became apparent that the
higher energy and larger beam size would make such a scheme too unwieldy.

\begin{figure}
\begin{centering}
\epsfig{figure=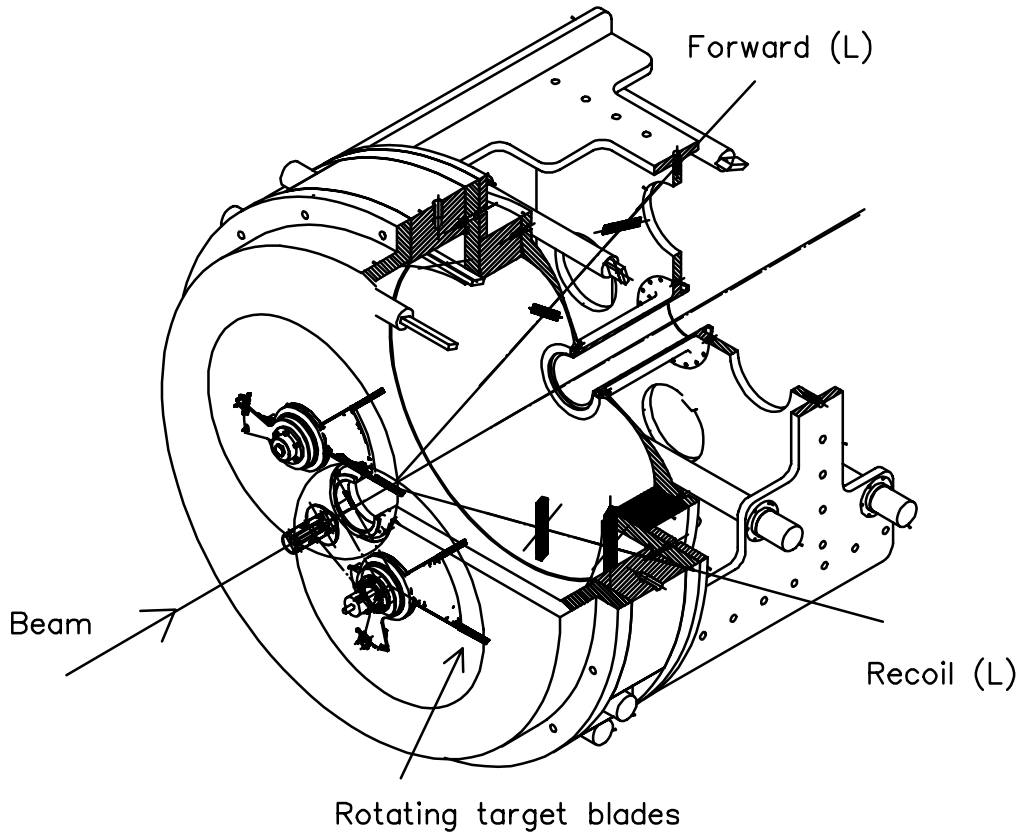,angle=0}
 \caption{General schematic view of a PPM.  The forward and recoil
paths for one arm are shown with the scintillators indicated as dark
volumes.  The paths originate from the plane in which the blades lie.}
\label{fig1}
\end{centering}
\end{figure}

The mechanics of the present detector have been described in \cite{sou:94},
though there have been modifications since then which will be
explained below.  
The device is a four-branch polarimeter whose target
consists of two wheels that can drive strips (`blades') of $CH_{2}$ (two on each
wheel) through the beam at a speed locked to the experiment cycle time.
It is shown in Figure~\ref{fig1}.
Two blades per wheel were chosen as an optimal compromise between
polarization measuring time and $A_{z}$ (i.e., experimental determination
of the helicity dependent asymmetry of the beam transmission through an
$LH_{2}$ target) measuring time.

\begin{figure}
\begin{centering}
\epsfig{figure=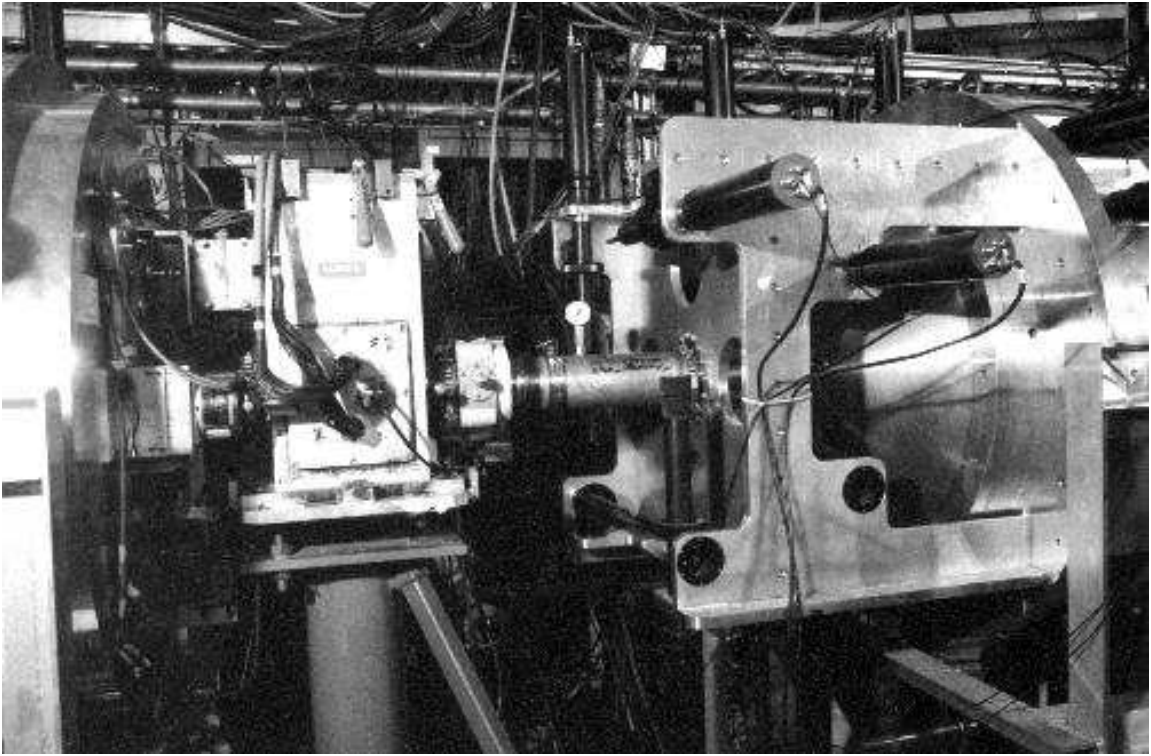,height=10cm, angle=0}
%\vspace*{10.cm}
 \caption{A section of the E497 experiment along TRIUMF beamline 4A/2
showing the whole of the upstream PPM on the right, the rear of the 
downstream PPM on the left, and, between them, a beam position 
monitor [19]. 
Several scintillators, light guides, and PMT housings can clearly be seen
on the upstream PPM.  }
\label{fig2}
\end{centering}
\end{figure}

Two of these detectors are mounted in the experimental beamline to
allow for extrapolation of the polarization profiles to the target
location.  They are shown in Figure~\ref{fig2}.  They are upstream of the
Transverse field Ionization Chambers (TRIC's) that sandwich the 
$LH_{2}$ target.  The TRIC signals (proportional to the beam current)
determine the parity violating longitudinal analyzing power that is
the observable of interest.
% Photograph for b/w reproduction
%\cite{sek:1}.

\subsection{PPM Detectors}
\begin{figure}
\begin{centering}
\epsfig{figure=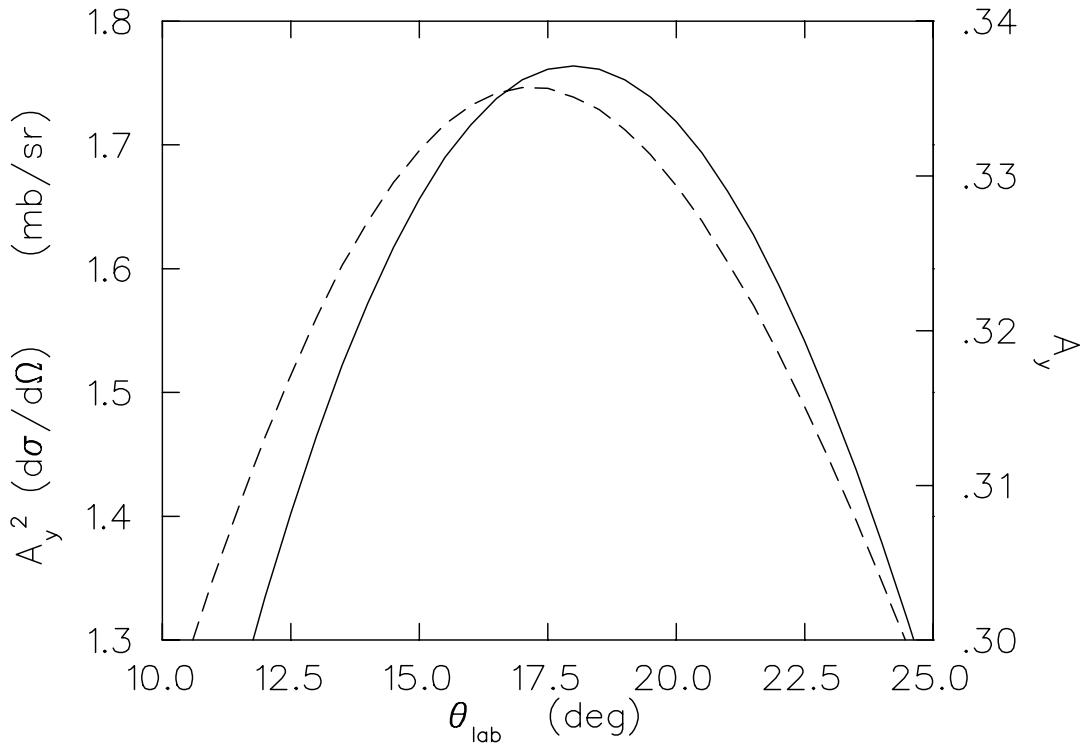,width=0.7\textwidth,angle=90}
 \caption{Analyzing power (solid line) and figure of merit (dashed line)
as a function of lab angle for $p \, + \, p$ scattering as determined from
Ref. \cite{SAIDx}.  
The peak in figure of merit is ideal for optimal statistical
error and the peak in analyzing power is optimal (flattest response
as a function of target position) for
systematic errors.}
\label{fig3}
\end{centering}
\end{figure}

Each branch consists of a forward arm of two scintillators at $17.5^{ \circ }$
from the axis and a recoil arm.  The angle of $17.5^{ \circ }$ was chosen as
a reasonable compromise near the $p-p$ analyzing power maximum over the energy
range at which parity violation may be investigated at TRIUMF, see 
Figure~\ref{fig3}.
The figure of merit for a polarimeter can be defined as:\\
\begin{equation}
A_{t}^{ \! 2}( \theta ) \frac{d \sigma }{d \Omega } ( \theta ).
\end{equation}
This is also shown in Figure~\ref{fig3}.

%			- A(theta) & figure of merit (diagram)

\begin{figure}
\begin{centering}
\epsfig{figure=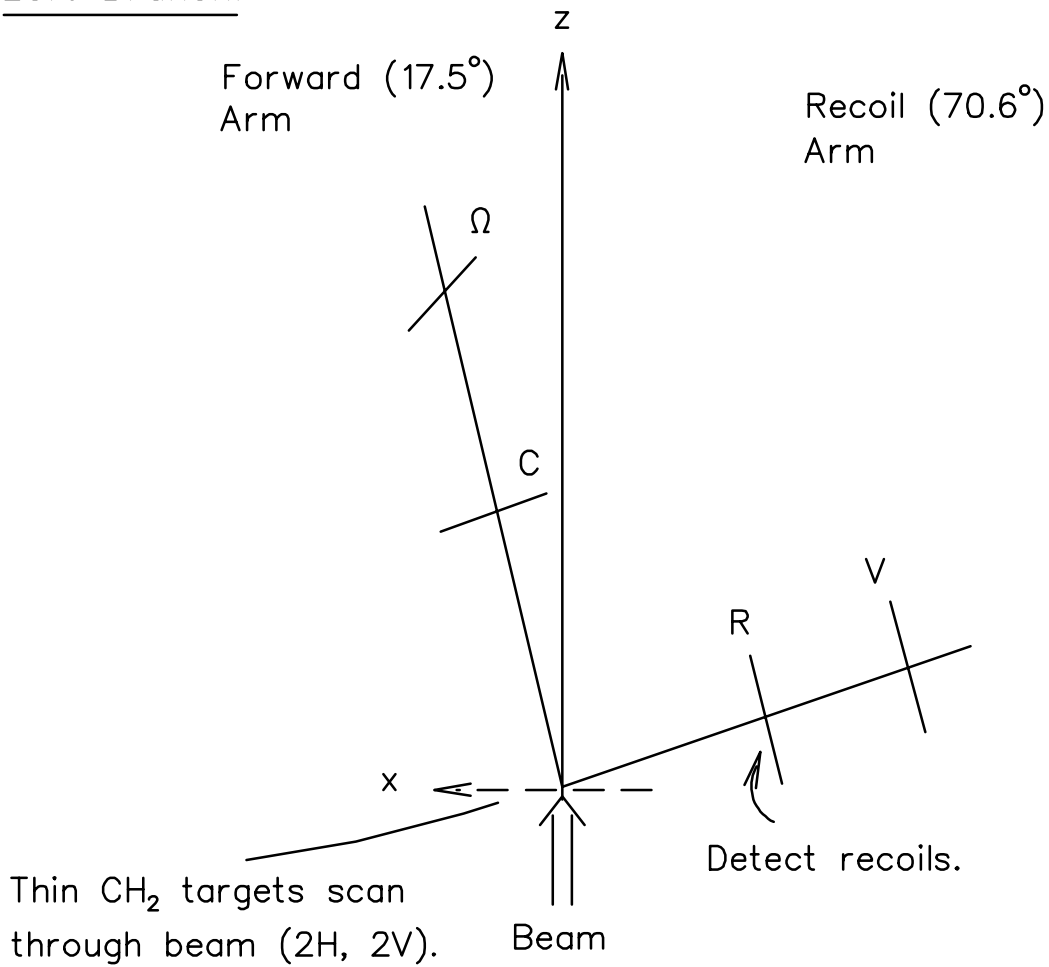,angle=0}
 \caption{Schematic diagram of a single branch of one of the PPM's.}
\label{fig4}
\end{centering}
\end{figure}

The forward arm consists of two scintillators.  The solid-angle of
acceptance for scattered protons is defined by a rotated counter
($ \Omega $) whose angle of rotation along its axis perpendicular
to the scattering plane is chosen to cancel the effect of the change
in $p-p$ scattering cross-section and detector geometry with target
blade position \cite{gree:dna}.  
At 223 MeV, this rotation angle is determined to be $49^{ \circ} $ with
respect to the plane perpendicular to the nominal $17.5^{ \circ }$
center-line of the scattered protons, in the direction as shown in Figure 4.
Between the $ \Omega $-counter and the
target plane is a counter ($C$) whose function is to determine that the
scattered protons are collinear with the target.
The recoil arm is at $ 70.6^{ \circ }$ at which the recoil protons from
$p-p$ scattering will be stopped in the front ($R$) counter or in a 1.6 mm thick
aluminum shield immediately behind it.  Protons from other sources that
are too penetrating will pass through and hit the veto ($V$) counter
immediately behind.  The location and dimensions of each counter are recorded
in Table~\ref{tabI}.
A schematic of a single branch lay-out is given in Figure~\ref{fig4}.

\begin{table}[t]
\begin{centering}
 \caption{PPM scintillator counter dimensions.  Distances are from the center
of the target plane. }
\begin{tabular}{|cccccc|}   \hline
Counter & Height & Width & Thickness & Distance & Arm \\
 & (mm) & (mm) & (mm) & (mm) & \\ \hline 
C & 37.5  & 37.5  & 6.4  & 600.2  & Forward \\ 
$ \Omega $ & 28.5  &  46.0$^{*}$ & 6.4  & 900.0  & Forward \\ 
R & 120.2  & 22.5  & 6.4  & 104.4  & Recoil \\ 
V & 156.9  & 30.0  & 6.4  & 151.7  & Recoil \\ \hline
\end{tabular}
\\
$^{*}$Counter is rotated at 49$^{ \circ }$ with suitably beveled edges.\\
\label{tabI}
\end{centering}
\end{table}

%                       - counter arrangements (diagram)

Each scintillator\footnote[3]{BC-404; Bicron; 12345 Kinsman Rd., Newbury, OH,
U.S.A., 44065}
was attached to a light-pipe viewed by a two-inch RCA 8575\footnote[4]{RCA
Corp.   }
photomultiplier tube.  The TRIUMF-built bases
% Do we have a reference to this?  GHC? or who? IEEE?
were equipped with zener-diodes on the first three dynodes and the voltages 
had to be carefully adjusted due to the high rates. 
The front arm counters, $C$ and $ \Omega $, of each branch were
mounted externally with the protons
passing through a 3.2 mm (at $17.5^{ \circ }$) thick spun-aluminum shell, 
at a distance of 470 mm
from the target, into air.  The recoil arm counters, $R$ and $V$, were mounted
internally with a vacuum seal along each scintillator light pipe.  
The external counters and light pipes were wrapped with 
aluminized mylar and tape
to keep out the ambient lighting; the internal counters were wrapped
with a light-tight aluminum foil only and their external sections
of light pipe were wrapped as the external counters.

%			- blade targets, motor drive, timing signals

\subsection{PPM Targets}
\begin{figure}
\begin{centering}
\epsfig{figure=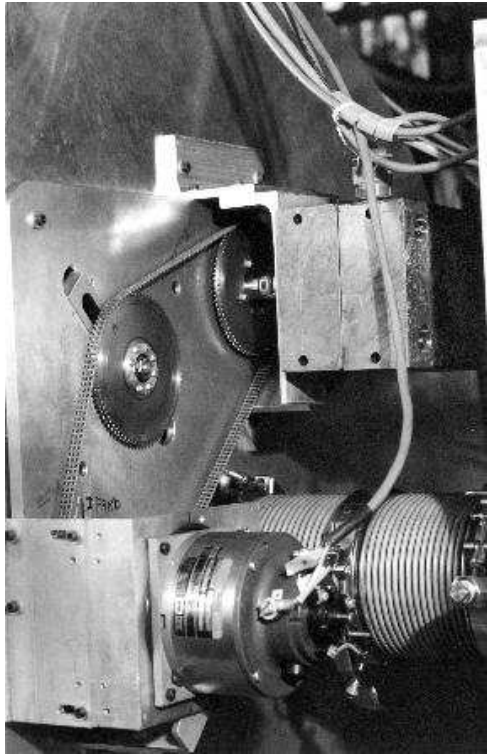,height=10cm, angle=0}
%\vspace*{10cm}
 \caption{ A view of the rear of a PPM showing the external belt drive,
the stepping motor at bottom, and the shaft encoder shielding at top.}
\label{fig5}
\end{centering}
\end{figure}

The mounting of the target blades and the drive arrangement is shown
in Figure~\ref{fig5}.  The wheel pivots are 215 mm from the beam centre.
Each arm holds two targets to better balance the statistics of the PPM
with the experiment statistics.  This gives two $x$ scans and two $y$
scans per PPM, a total of eight.  Each scan occurs during one spin-state
of an eight state cycle.  The direction of the spin in each state is
defined by the eight state cycle which can be $(+ \, - \, - \, + \, - \, + \, 
+ \, -)$ or its complement.  The initial state of each cycle is chosen
according to the same $(+ \, - \, - \, + \, - \, + \, + \, -)$ pattern,
making up a 64-state `supercycle'.  The initial spin direction of each 
supercycle is chosen randomly.
This timing sequence is shown in Figure~\ref{fig6}.
% timing signals

Each blade target is 1.6 mm wide, 5 mm along the beam, and 85 mm high
(past its holder-clamp)
and is machine cut from sheets of high density polyethylene.
As each blade passes through the beam, a proton scattered in a plane 
containing the direction of motion of the blade is observed
in one of the two forward arms, left-right (horizontal motion) or bottom-top
(vertical motion), and the corresponding recoil proton from free $p-p$ 
scattering is observed in the recoil arm on the opposite side.
Protons scattered in a plane perpendicular to the direction of
motion of the blade (i.e., those that would give $P_{x}(x)$ and $P_{y}(y)$)
are not recorded as their recoil protons would in many cases be stopped
or severely multiple-scattered in the target.  

The target blades are driven through the beam by a D.C. servo-motor/tach\-o\-meter
unit\footnote[5]{Electro-Craft Corporation; 1600 Second St. So., Hopkins,
MN, U.S.A. 55343} salvaged from an old reel-to-reel tape drive.
The two wheels are connected by a timing belt
that is mounted external to the PPM
housing.  This was done because it was proven necessary to ensure 
proper cooling.  The power to the wheels is transmitted through
ferrofluidically-sealed shafts\footnote[6]{Ferrofluidics Corp.; 40 Simon St.,
Nashua, NH 03061, U.S.A.}.
The read-out of the shaft position was done through a shaft
encoder\footnote[7]{Type H25D; BEI Sensors and Motion Systems Co., 
Industrial Encoder Division; 7230 Hollister Ave., Goleta, CA, U.S.A. 
93117-2891}.  It was found necessary to shield this encoder and 
switch to a rad-hard version as radiation damage caused failure
after the first few weeks of running.  This has not been a problem since.

In addition, with the blades turned off and parked out of the beam, it
is possible to insert a fixed target of $CH_{2}$ some 0.2 mm thick.
This target has a very thin film of aluminum evaporated on the surface
to prevent charging, and is mounted in a circular aluminum frame
100 mm in diameter.
This allows a rapid determination of $P_{y}$ and $P_{x}$ in the
parity beamline, useful for initial tuning of the solenoids that
provide longitudinal polarization.

\subsection{Synchronization and Control}
The PPM's rotate at five revolutions per second and are adjusted for
$180^{ \circ }$ angular mismatch.  A full 200 ms cycle compromises
eight blade passages with 25 ms between passages.  The synchronization 
of the PPM's, as well as the maintenance of the rotation speed, is accomplished 
by an application of electronic gearing.  Each PPM is equipped with a 2500 
line incremental shaft encoder and DC brushed servo motor.

\begin{figure}
\begin{centering}
\epsfig{figure=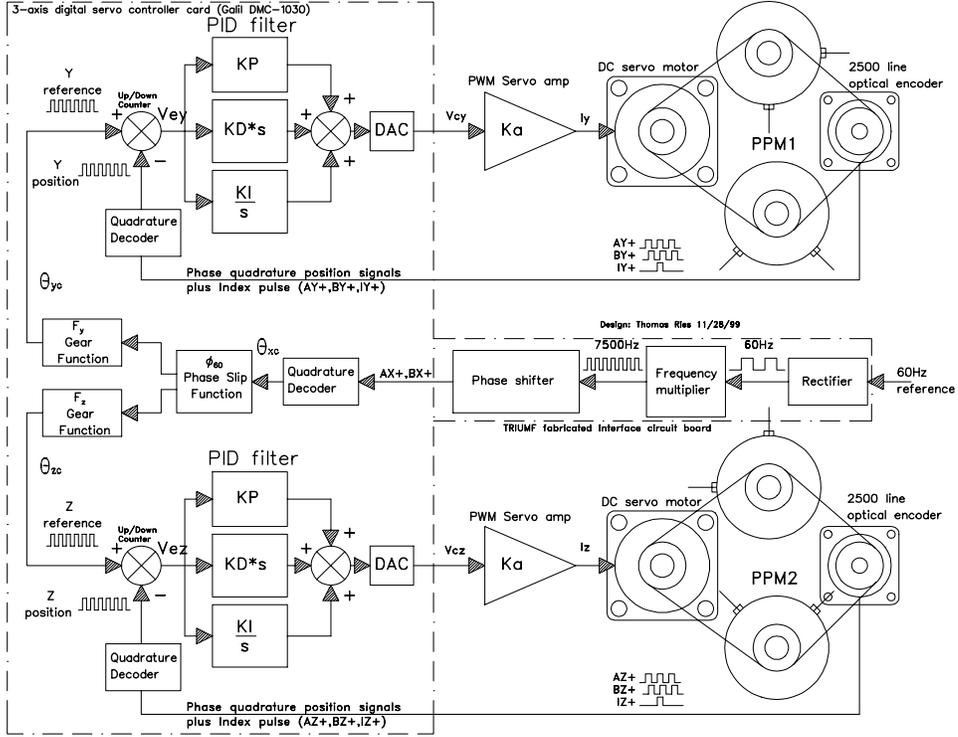,width=0.7\textwidth,angle=90}
 \caption{Schematic of the PPM control system.  The $Y$ and $Z$ references
and positions are input in quadrature counts, $+$ and $-$ respectively to
the Up/Down counter, whose output is fed to the PID filter.  The PID filters' 
outputs are fed
to Pulse Width Modulated (PWM) switching 20 kHz amplifiers in current/torque
mode.  The amplifier output runs the DC servo motor as discussed in section 2.2.
The optical encoders provide the position signals.  The reference circuit,
adjusted by $ \phi _{60} $ is used to provide the reference signals.
}
\label{gal}
\end{centering}
\end{figure}

The motors are controlled by a Galil DMC1030\footnote[8]{Galil Motion Control,
Inc.; 203 Ravendale Drive, Mountain View, CA, U.S.A. 94043-5216} 3-axis PC ISA 
bus based
digital servo motion control card.  A functional block diagram of
the control system is shown in Figure~\ref{gal}.  A reference 60 Hz
square wave signal is generated from the 60 Hz AC line which
has a frequency regulation of 0.06\%.  This signal is frequency multiplied
by a factor of 125 and phase locked to the 60 Hz line via a voltage
controlled oscillator (VCO) feedback 
regulator circuit.  The resulting 7500 Hz is phase shifted to produce a 
double phase quadrature signal.  This is directed to the $x$-axis encoder
input of the DMC and represents the master axis signal which the slave
axes, $y$ and $z$, are commanded to follow through the gear function
ratios, $F_{y}$ and $F_{z}$.  The phase slip function factor, $ \phi _{60}$,
utilizes machine round off error which comes from the fact that 2500 is 
not evenly divisible by 60, so that at 5 Hz the right amount of phase slip 
relative to 60 Hz is obtained.  This means that rotator speeds of 3 Hz,
6 Hz, 9 Hz, etc., can be set precisely to zero phase slip, while speeds
in between cannot (unless the encoder resolution were changed to 3000 lines
per turn).  Normally, the phase slip is set to one 60 Hz cycle in 20 minutes.
At 5 Hz, the encoder frequency is perfectly suited for this application.
However, the factors $ \phi _{60}$, $F_{y}$, and $F_{z}$ can be configured 
interactively by the user from the windows graphical user interface (GUI) 
at any time.

In normal operation the gearing is set for 1:1 on both PPM's and the
phase difference between the two PPM's is set for $180^{ \circ }$.
To compensate for small mechanical misalignment in the mechanisms, a fine
phase adjustment is made so that the actual blade passages through the
beam (between the two PPM's) are exactly $180^{ \circ }$ apart.
The gear ratios $F_{y}$ and $F_{z}$ modify the output signal, $ \theta _{c}'$
from the $x$-axis phase slip function and produces command frequency references,
$ \theta _{yc}$ and $ \theta _{zc}$, so the PPM speed is correctly calibrated,
as required by the user, based on the 60 Hz line signal.  During standard
use $F_{y} \, = \, F_{z}$, which means that the two PPM's are phase locked
to run at the same speed with zero relative phase slip.

Measurements with a digital oscilloscope showed that during rotation at
5 Hz the servo loops kept the two PPM's within $ \pm 1$ encoder tick (i.e,
$ \pm 0.003$ radians) of each other.  The two reference signals, $ \theta
_{yc}$ and $ \theta _{zc}$, are treated by the microprocessor as
quadrature counts which are the Basic Length Unit at machine hardware
level.  This means that the phase synchronization and position accuracy
of the servo loop is four times greater than the line frequency of the encoder.
These signals are compared in an up/down counter against the encoder
feedback signal and the difference is used to produce an analog command 
voltage signal for the servo amplifier via a PID filter and DAC running
at a sampling rate of 1 kHz.  This produces the current to drive the motors.
The PID filter parameters, $KP$, $KD$, and $KI$, are the same for both
PPM's due to their similar plant dynamics and shaft torque resistances.
However, the stable operating region is very narrow due to the flexible
couplings and the large inertia mis-match between the motor armature and
the blade rotor mechanism (required due to space constraints). The aim is
to increase $KP$ in order to minimize the phase lock and position error,
but not high enough to make the static gain loop unstable.  To help
stabilize the latter, $KD$ is increased high enough to damp out the
low frequency instabilities, but not high enough to destabilize the
derivative loop gain.  $KI$ is set to zero in order not to induce low
frequency oscillations into the loop due to the high load inertia.
An interesting aspect to Figure~\ref{gal} is that the distance between the
PPM/Servo-Amp units at beam line level and the DMC controller is over
150 meters.  This is very unusual in servo control applications due to the
destabilizing effects of phase delay in the encoder and command signal
cables; but was required due to the radiation environment.

For control measurements, it is possible to run a single PPM or to
park a PPM's blades at a specific angle.

\section{Signal Processing and Data Acquisition}
\begin{figure}
\begin{centering}
\epsfig{figure=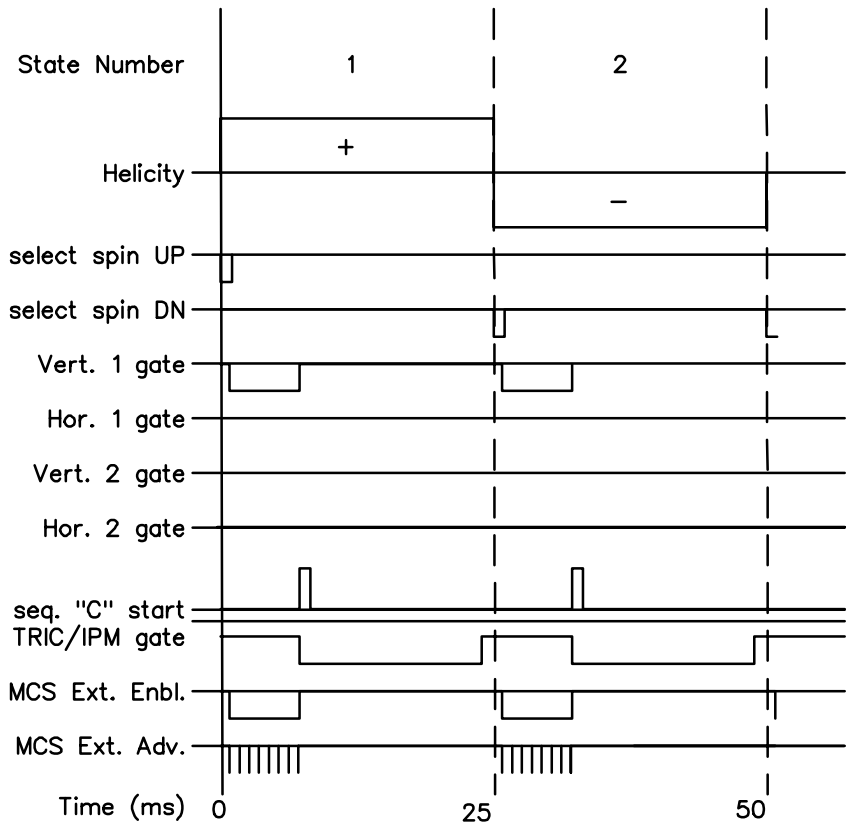,angle=90}
 \caption{The data collection cycle.  
Shown are the first two states of the eight-state cycle, which can be
$( \, + \, - \, - \, + \, - \, + \, + \, -)$ or its complement.
During each state of the cycle, one of the eight PPM blades passes
through the beam.  For the two states shown here, the two vertically
scanning blades of PPM1 would scan during the `Vert. 1 gate' intervals.
While the blade is passing through the beam
the appropriate electronics and MCS channels are gated/enabled and
the MCS channel is advanced by a signal from the shaft encoder (see 
Fig.~\ref{fig7}).  The PPM data is read out while the other equipment
(TRIC/IPM) data collection is enabled.  The spin state selection and
sequencer start are controlled by the front-end processor.}
\label{fig6}
\end{centering}
\end{figure}

The PPM data collection is an integral part of the experimental data
collection cycle.  The PPM blade scans are carefully synchronized not
only to each other, but are used to drive the
polarized source spin flip cycle and the signal integration gates on the two
transverse ion-chambers that bracket the target and whose helicity-dependent
output constitutes the parity-violating signal of the experiment.  
Each shaft encoder pulse forms the time-base for the experiment
as an input to a timing and sequence module\footnote[9]{Model 221; Jorway
Corp.; 27 Bond St., Westbury, NY, U.S.A. 11590}.  A
schematic diagram of the data collection cycle is shown in Figure~\ref{fig6}.

\subsection{Electronics}
\begin{figure}
\begin{centering}
\epsfig{figure=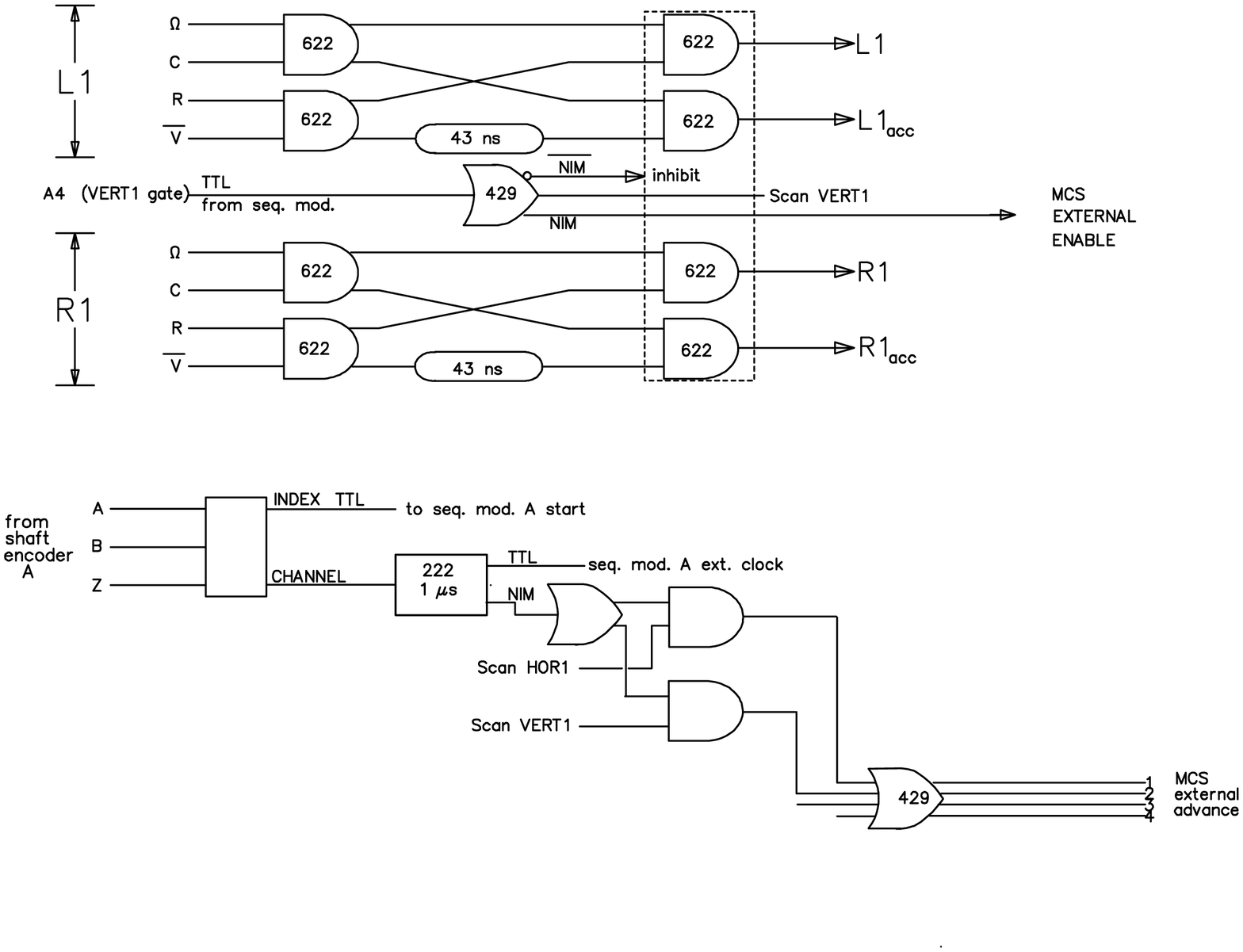,width=0.7\textwidth,angle=90}
 \caption{Schematic of the PPM electronics, two branches of one four-branch
PPM are shown.  The discriminated signals from each counter (10 ns width)
are timed in to form the logic coincidence for each branch, {\it Ln, Rn,
Bn, Tn,} and their corresponding delayed coincidence (accidental), {\it 
Ln$_{acc}$, Rn$_{acc}$, Bn$_{acc}$, Tn$_{acc}$}; $n \, = \, 1,2$ 
labelling either one of the two PPM's.  The four signals for any PPM plane 
(horizontal {\it [Ln, Rn, Ln$_{acc}$, Rn$_{acc}$]} or vertical 
{\it [Bn, Tn, Bn$_{acc}$, Tn$_{acc}$]}) are OR'ed together with the other
planes and presented to the four MCS units, suitably gated for the appropriate 
blade as explained in the caption to Figure 6. }
\label{fig7}
\end{centering}
\end{figure}

A schematic lay-out of the PPM electronics is given in Figure~\ref{fig7}.

The signals from the phototube bases were fed through an amplifier,
thus allowing the tubes to be run at lower voltages, important due
to the high singles rates, and into individual linear discriminators.
Each pair of forward arm signals, $C$ and $ \Omega $, were formed
into a logical coincidence, $C \cdot \Omega$, and each recoil
arm was formed into an anticoincidence, $R \cdot \bar{V}$.
These were timed together to form $(C \cdot \Omega) \cdot (R \cdot \bar{V})$
and $(C \cdot \Omega) \cdot (R \cdot \bar{V})_{ \mbox{del}}$ ({\it del} 
indicating that the signal has been delayed by one cyclotron RF period ---
43 ns), where the
first is the coincident signals, $L$, $R$, $B$, and $T$, and the latter
are their corresponding accidentals, $L_{a}$, $R_{a}$, $B_{a}$, and $T_{a}$.
These signals are grouped together in common modules for $L-R$ and
$B-T$ and for the two PPMs, which can be inhibited by the timing
sequence.  This allows a fan-in of the signals, for example ${L1}$,
${B1}$, ${L2}$, and ${B2}$, together and they are then presented to the
same scaler input, as their respective blades are never in the beam at the
same time.  

The scanning scalers and memory modules\footnote[10]{3521A, MM8206A; LeCroy Research
Systems; 700 Chestnut Ridge Road, Chestnut Ridge, NY, U.S.A. 10977-6499}
read in the data in synch with a clock signal.  As
each blade moves through the beam the scaler advances through a sequence
of channels that are related to the position of the blade.
%role of encoder?
The data is then read out through a routine running in a dedicated
processor\footnote[11]{Starburst J11; Creative Electronic Systems; 70 route
de Pont Butin, 1213 Petit-Lancy 1, Switzerland} that stores
the results in memory according to the timing sequence, {\it e.g.},
$L_{I}$, $B_{I}$, $L_{II}$, and $B_{II}$, and the spin-state.
Thus there is a requirement for only four such scalers and memory modules
to record two true and two accidental signals per blade.  This allows
many of the more crucial experimental modules to reside in a single
crate, important for the in-crate control through the Starburst and
timing sequences.
The status of each spin state ($+$ or $-$ helicity) is nowhere introduced
as a gating signal to any of the hardware; thus avoiding undesirable
cross-talk which might lead to a helicity-dependent electronics effect.
Rather, the spin state status of the initial state (state 1, see Fig. 7)
in a pattern of eight states is separately reported, as a frequency
modulated/encrypted signal, to the computer.

\subsection{Data Acquisition}
The PPM information was read out of the front-end processor as a separate
event, there being separate events for the TRIC and other monitor information.
This allowed the PPM information to be transferred to the main
data acquisition computer\footnote[12]{VAXstation 3200; Digital Equipment Corp.;
Maynard, MA, U.S.A.} while other data was being
collected, and {\it vice versa}.  The data was then written to tape and
made available to other processors for on-line analysis and monitoring.
The last was especially important for the PPM data as it allowed us
to monitor both transverse polarization components and the first moments
of polarization on a run-by-run (approximately hourly) basis.  If these
observables became excessively large, then the beam-line solenoids or
other cyclotron parameters were tuned to reduce them.  As PPM information
was available in each data buffer, each buffer (200 ms of data) could be
analyzed separately and bundled as seemed appropriate for a
regression analysis.  A first analysis of such a kind was done in
a semi-online manner so that more sophisticated monitoring of the 
experiment could be carried out.

\section{Results}
The PPM counter's were run at a comparatively high rate.  Table~\ref{tabII} 
shows the peak
rates for both singles in each individual counter and the coincidence rate
at a beam current of 200 nA and a size of 5 mm.  At this current, first
order accidentals (forward arm accidentally in coincidence with the
recoils arm) were typically 35\% of the 
$ (C \cdot \Omega) \cdot (R \cdot \bar{V} ) $ coincidence rate.

Two higher order accidentals were examined: (1) $ C \cdot (R \cdot \bar{V} ) $
with an accidental hit in the $ \Omega $ counter; and 
(2) $(C \cdot \Omega) \cdot 
R$ with an accidental hit in the $V$ counter.  The first are `near' events
in the sense that they are close to the acceptance of the PPM with a similar
(very slightly lower, see Fig.~\ref{fig3}) analyzing power; and the second are `stolen' 
events in that they would have 
been accepted as true events but for the accidental veto.  Both might result
in errors in the measured intrinsic first moment of polarization coupled
to a helicity correlated change in the beam intensity (otherwise, they just
tend to pull down the average analyzing power slightly).  The
effects were measured by taking data with a 43 ns (one RF period for the
TRIUMF cyclotron) delay in the $ \Omega $ counter for case (1), and a
43 ns delay in the $V$ counter for case (2).  For an assumed helicity
dependent variation of current, $ \frac{ \Delta I}{I} \, = \, 10^{-5}$,
the change in the first moment due to case (1) was $ 1.2 \pm 0.2 \times
10^{-3} \mu \mbox{m}$, and for case (2) was $ 0.0 \pm 0.2 \times 10^{-3}
\mu \mbox{m}$.  As these would result in false terms to $A_{z}$ of the
order of $10^{-11}$, they were inconsequential for the experiment.

\begin{table}[t]
\begin{centering}
 \caption{Singles and coincidence rates in PPM detectors. }
\begin{tabular}{|cc|} \hline
Detector & Peak Singles Rate (MHz) \\ \hline  
C & 3.1 \\
$ \Omega $ & 0.8 \\
R & 3.0 \\
V & 1.5 \\ \hline
Coincidence & Peak Rate (kHz) \\ \hline
$ (C \cdot \Omega) \cdot (R \cdot \bar{V} ) $ & 110. \\
$ (C \cdot \Omega) \cdot (R \cdot \bar{V} \, + \, 43 \mbox{ns} ) $ 
& 38. \\ \hline
\end{tabular}
\label{tabII}
\end{centering}
\end{table}

Tests were also done with carbon blades replacing the usual $CH_{2}$
blades in the polarimeter.  These indicated that 1\% of the true events
in the PPM came from the carbon in the $CH_{2}$ blades ($^{12}C(p,2p)X$, etc.).
This had a very small contribution to the effective analyzing power.

\begin{figure}
\begin{centering}
\epsfig{figure=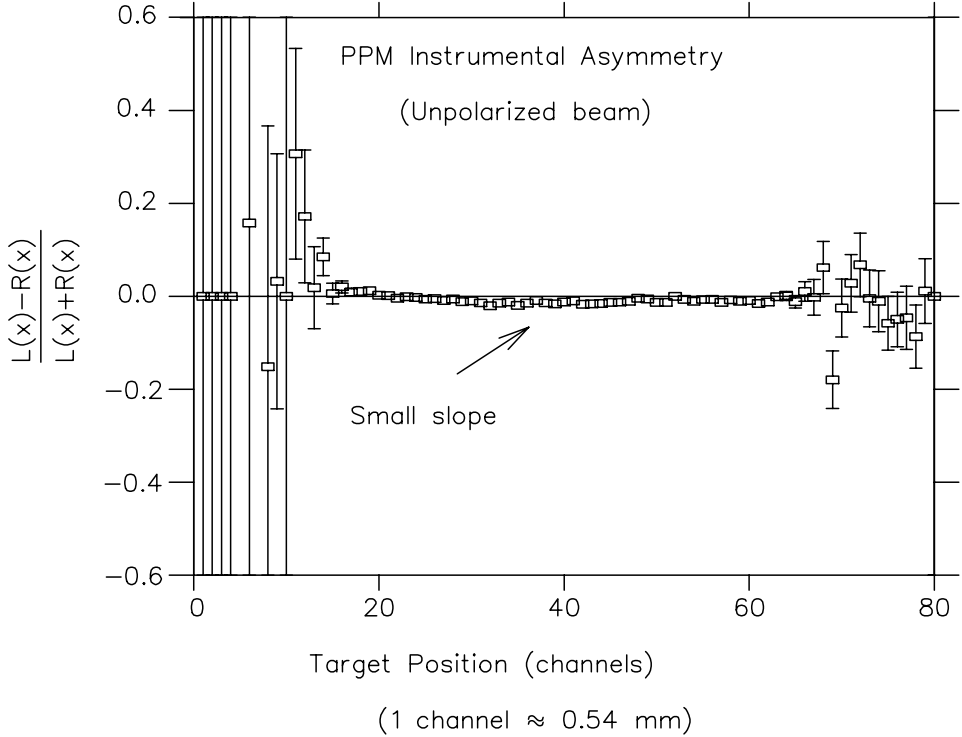,width=0.7\textwidth,angle=90}
 \caption{Instrumental asymmetry as a function of target (blade) position.}
\label{fig8}
\end{centering}
\end{figure}

As the data collection involved spin off periods interspersed with the
polarized beam, it was possible to monitor the PPM's response (instrumental
asymmetry) to zero polarization (ideally what we would like to see in
the experiment with a perfect longitudinally polarized beam).
An instrumental asymmetry as a function of blade position plot is presented
in Figure~\ref{fig8}.  
It was found that the slope of the instrumental asymmetry
was strongly dependent on the divergence or convergence of the beam, as
then the angle of incidence is position dependent and the scattering angle
dependence on position is different from the assumption of a parallel
beam.  Note that the requirement for the experiment is for as parallel
a beam (i.e., very weakly focussed at the target) as reasonably achievable.

\begin{figure}
\begin{centering}
\epsfig{figure=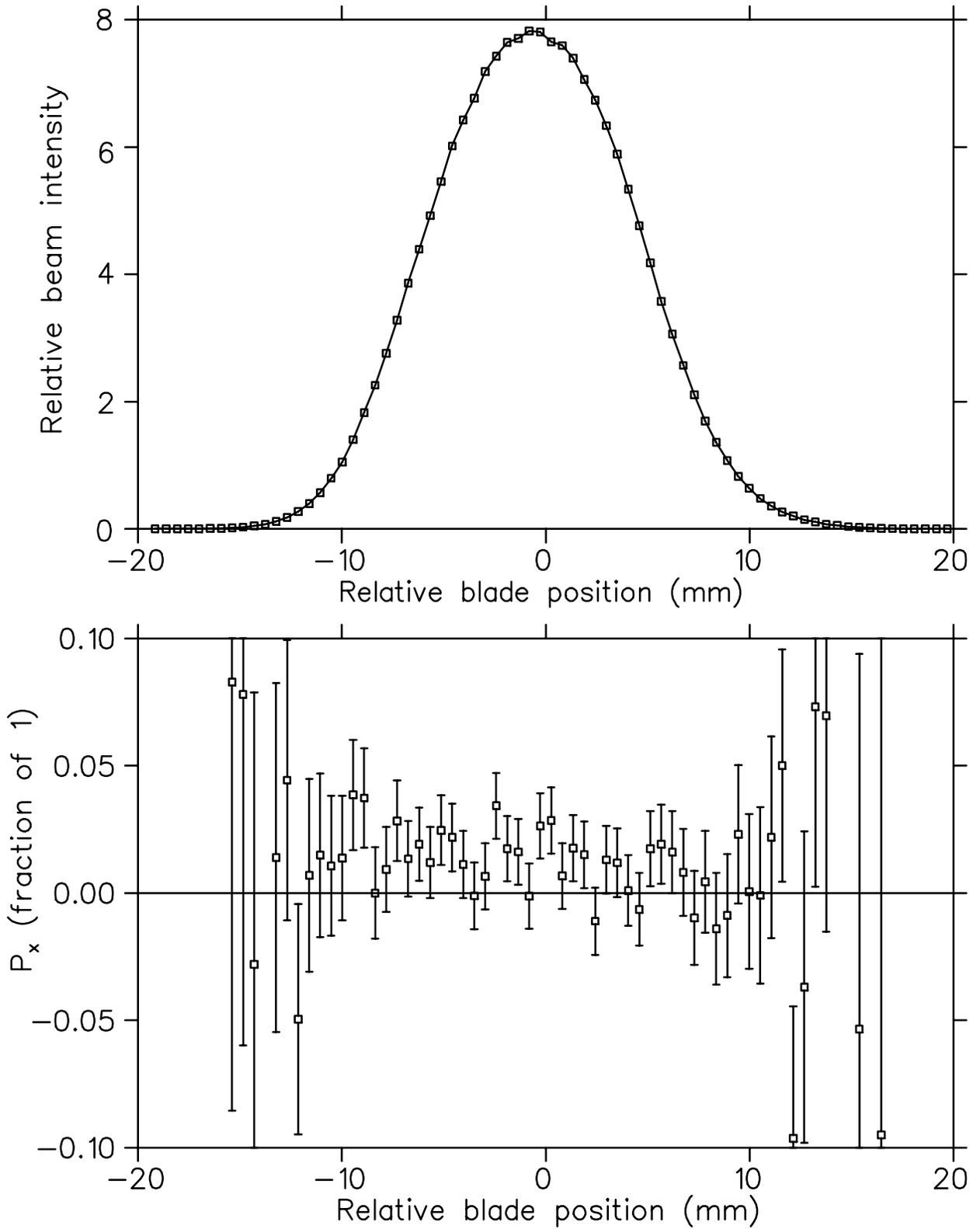,angle=0}
 \caption{A beam profile (top) and polarization profile (bottom) for a
longitudinally polarized beam.  In this example $<P_{x}>$ is obviously
non-zero (about 1\%).}
\label{fig9}
\end{centering}
\end{figure}

Figure~\ref{fig9} shows a helicity-correlated polarization profile measured
by the upstream PPM with 200 nA beam and a beam size ($ \sigma $) of 5.0 mm.
Under those conditions, each PPM measures an average $<P_{x}>$ and $<P_{y}>$
to $ \pm $0.002 and $<xP_{y}>$ and $<yP_{x}>$ to $ \pm $7 $ \mu $m in
one hour.

The effective analyzing power as a function of blade position is
determined by moving the beam across the range of the blade sweep
with the beam transversely polarized.  
Absolute calibration was done by comparing the integrated result
to the existing IBP.

\section{Conclusions}
The PPM rotation control system has worked very well.  It is convenient
to use and normally maintains PPM synchronization to $ \pm $ one shaft encoder
line.

The PPM's have been successfully used throughout the TRIUMF Parity
experiment (E497).  For three runs (not the full data set) of data
taken in 1997, 1998 and 1999 (about four months) consisting of about 
240 hours of TRIC data (the actual parity violation measurement) collection:
The `false' parity violating analyzing power
($A_{z}$) derived from transverse components of polarization coupled with
a displacement from the ideal instrumental symmetry axis has been
measured as $( 0.02 \pm 0.01) \times 10^{-7} $.  The false $A_{z}$ 
derived from the first moments of polarization has been measured as 
$( 0.72 \pm 0.19) \times 10^{-7} $.
This confirms the expectation that the latter is a large (indeed, so far, the
largest) correction.  It is also the largest contribution to the E497 error.
Improved PPM error to total error could be achieved by changing the
number of targets (and thus the ratio of PPM data collection time to
TRIC data collection time) or seeking some other means of rapidly and accurately
measuring the polarization profiles.

%	    REFERENCES

\end{document}